\documentclass[aps,preprint,amssymb,showpacs,eqsecnum,tightenlines]{revtex4}

\usepackage{amsmath,amsfonts,amssymb,amsthm}

\renewcommand{\(}{\left(}
\renewcommand{\)}{\right)}
\renewcommand{\[}{\left[}
\renewcommand{\]}{\right]}

\newcommand{\beq}{\begin{equation}}
\newcommand{\eeq}{\end{equation}}
\newcommand{\beqa}{\begin{eqnarray}}
\newcommand{\eeqa}{\end{eqnarray}}
\newcommand{\nol}{\nonumber}

\begin{document}

\title{A conformally coupled massive scalar field in the de~Sitter expanding universe with the mass term treated as a perturbation}

\author{Atsushi HIGUCHI$^1$ and LEE Yen Cheong$^2$}

\affiliation{Department of Mathematics, University of York,
Heslington, York YO10 5DD, United Kingdom\\ email:
$^1$ah28@york.ac.uk, $^2$yl538@york.ac.uk}

\date{16 May, 2009}

\begin{abstract}
We study conformally-coupled massive scalar field theory with the mass term treated as a perturbation in the expanding half of de Sitter spacetime as a toy model for comparing various perturbative formalisms.
We point out that the in-out perturbation theory breaks down due to infrared divergences coming from the infinite future.  We then clarify the relation between the free-theory vacuum and the true vacuum using the Bogolubov transformation and show that the discrepancy between the free-theory out-vacuum and the true vacuum causes apparent pair creation of free-theory particles and makes the in-out two-point function differ from the true two-point function.  We also
identify the infinite Bogolubov coefficients as the cause of infrared divergences.
We then examine two alternative approaches: the Euclidean and in-in formalisms.  We verify that there are no infrared divergences in perturbation theory with the mass term treated as a perturbation in either of these approaches and that the two-point function of massive scalar field is reproduced correctly by these methods.
\end{abstract}


\maketitle

\section{Introduction}

It is well known that lack of a global timelike Killing vector field in general leads
to non-uniqueness of the vacuum state~\cite{Birrell:1982ix}. (See Refs.~\cite{Parker:1968mv,Parker:1969au,Sexl:1969ix} for early work on this subject).  However, although de~Sitter spacetime lacks a global timelike Killing vector field, there is a unique `physically acceptable' de~Sitter-invariant vacuum state called the Euclidean or Bunch-Davies vacuum for free scalar field theory due to high symmetry of this spacetime~\cite{Gibbons:1977mu,Bunch:1978yq} (unless the field is minimally coupled and massless~\cite{Ford:1977in,Ratra:1984yq,Allen:1985ux}).  In this vacuum state an Unruh-DeWitt particle detector~\cite{Unruh:1976db,DeWitt:1980hx} would click, but the eigenvalue of the number operator corresponding to the annihilation and creation operators for the Fock space remains zero at all times.
Thus, there are no difficulties in free scalar field theory in de~Sitter spacetime (except for the minimally-coupled massless case), but the usual in-out perturbation theory in interacting field theory tends to suffer infrared divergences. (These
divergences, which occur in spacetime integrals,
are due to the exponential expansion of the space and are different from the
usual infrared divergences in the momentum space.  They are sometimes called ``superexpansionary divergences''~\cite{Sasaki:1992ux}, but since this term is not very commonly used, we have opted for the term ``infrared divergences''.)
For example, it has been shown that the four-point function in $\varphi^4$ theory is infrared divergent in the expanding half of de~Sitter spacetime, i.e.\ de~Sitter expanding universe,
if the mass is smaller than a certain value~\cite{Sasaki:1992ux}.  As the authors of this work point out, these  infrared divergences indicate that the in-out formalism is inadequate in de~Sitter spacetime rather than that the theory itself is problematic.
To clarify this point it will be useful to analyse an exactly soluble model in de~Sitter expanding universe using the in-out perturbation theory and examine how this theory fails.  In this paper we study the in-out perturbation theory for a free conformally-coupled massive scalar field theory in this spacetime, which is, of course, exactly soluble, with the mass term treated as a perturbation, which we call `our model' throughout this paper.

To understand why the in-out perturbation theory can fail in de~Sitter spacetime it is useful to recall how it is justified in Minkowski spacetime (see, e.g.~Ref.~\cite{Peskin:1995ev}).  In the Minkowski case the total Hamiltonian $H$ of the interacting theory is normally time independent and bounded below.  It is assumed that the free-theory in-vacuum state $|0,{\rm in}\rangle$, i.e.\ the lowest eigenstate of the unperturbed Hamiltonian $H_0$ in the infinite past, has a nonzero overlap with the true vacuum state $|\Omega\rangle$, i.e. $\langle \Omega|0,{\rm in}\rangle \neq 0$.  These assumptions allow one to extract
the true vacuum state $|\Omega\rangle$ from the free-theory in-vacuum state $|0,\mbox{in}\rangle$
by taking the limit $t \to \infty(1-i\epsilon)$ in the expansion (in the Heisenberg picture)
\beq
e^{-iHt}|0,\mbox{in}\rangle = e^{-iE_0t}|\Omega\rangle \langle \Omega|0,\mbox{in}\rangle
+ \sum_n e^{-iE_n t}|n\rangle\langle n|0,\mbox{in}\rangle\,,
\eeq
where $E_0$ is the vacuum energy corresponding to the true vacuum $|\Omega\rangle$
and where $|n\rangle$ with $n\neq 0$ are the other states with energy
$E_n (> E_0)$, and similarly for the out-vacuum state $|0,\mbox{out}\rangle$.  This formula allows one
to express the time-ordered two-point function of a scalar field $\phi(x)$ in the theory, for example, as
\beq
\langle \Omega|T\phi(x)\phi(x')|\Omega\rangle =     \frac{\langle0,\mbox{out}|T\phi(x)\phi(x')|0,\mbox{in}\rangle}{\langle0,\mbox{out}|0,\mbox{in}\rangle} \,.
\label{AHPeskin}
\eeq
Then the right-hand side can be evaluated using time-dependent perturbation theory.
In de~Sitter expanding universe the Hamiltonian is time dependent, and therefore the formula (\ref{AHPeskin}) cannot be justified in the same way as in Minkowski spacetime.
In this paper we find that the right-hand side
does not give the true two-point function in `our model' of massive scalar field theory with
the mass term treated as a perturbation.

After examining how the in-out perturbation theory fails in `our model', we turn to two alternative methods: the Euclidean approach and the in-in formalism~\cite{Schwinger,Keldysh}.  It is well known that free scalar quantum field theory in de~Sitter spacetime with Hubble constant $H$ is equivalent to the thermal field theory with the Gibbons-Hawking temperature $H/2\pi$ if one only considers quantities inside the cosmological horizon~\cite{Gibbons:1977mu}.  (This conclusion can be extended to certain interacting field theories as well~\cite{Sewell:1982ap}.)  It is also well known that the thermal field theory inside the cosmological horizon is obtained by analytic continuation from Euclidean quantum field theory on the $4$-sphere ($S^4$) of radius $H^{-1}$.  We find that this Euclidean perturbation theory on $S^4$ recovers the correct two-point function in `our model'.

The in-in formalism has been advocated by many authors for studying quantum field theory in time-dependent background spacetime (see, e.g.\ \cite{Hajicek1978,Kay1980,Jordan:1986ug,Calzetta:1986ey}).
We find that this formalism, which reduces to finding the two-point correlation
function of the Heisenberg field
in the free-theory in-vacuum in `our model', reproduces massive scalar field theory correctly.

The rest of the paper is organized as follows.  In section~\ref{sec:in-out} we show that the in-out perturbation theory suffers infrared divergences in `our model'.  In section~\ref{sec:bogolubov} we calculate the time-dependent Bogolubov coefficients which relate the annihilation and creation operators for the massive scalar field and those for the massless scalar field, and use them to demonstrate that the true vacuum state is seen to have constant pair creation of free-theory particles relative to the free-theory vacuum state.  Then
in section~\ref{sec:bogolubov2} we use this Bogolubov transformation to show
that the exact
two-point function in the in-out formalism is infrared \emph{finite} but disagrees with the true two-point function for massive scalar field.  We clarify how infrared divergences emerge at first order in perturbation theory in spite of the finiteness of the exact in-out two-point function.
In sections~\ref{sec:Euclidean} and \ref{sec:in-in} we show that the Euclidean and in-in perturbation theories, respectively, yield the  two-point function correctly in `our model'.
We summarize our results in section~\ref{sec:remarks}.
In appendix~\ref{sec:emission} we use the standard perturbation theory to derive the free-theory particle creation rate found in section~\ref{sec:bogolubov} to first order in the squared mass.
In appendix~\ref{A} we explain why the in-out perturbation theory about a wrong-mass
works in Minkowski spacetime in spite of the fact that the corresponding Bogolubov transformation is nontrivial.
We use the metric signature $-+++$ and natural units $\hbar=c=1$ throughout this paper.


\section{Infrared divergences in the in-out perturbation theory}\label{sec:in-out}

Let us first recall how the time-ordered two-point function of a scalar field $\phi$
of mass $(M^2+m^2)^{1/2}$ can be constructed in the in-out
perturbation theory about a scalar field of mass $M$ in Minkowski spacetime.
The Feynman two-point function for the scalar field with mass $M$
is given by
\beq
    D_F(x,x') = -i\int\frac{d^4p}{(2\pi)^4}\frac{1}{p^2+M^2-i\epsilon}e^{ip\cdot(x-x')} \,.  \label{momexp}
\eeq
If we introduce a mass term, $-\frac{1}{2} m^2\phi^2$, in the Lagrangian as an interaction term, thus increasing
the squared mass from $M^2$ to $M^2 + m^2$,
the full Feynman two-point function in the usual in-out perturbation theory is
\beqa
    \Delta_F(x,x')
    & = & D_F(x,x')
    + \sum_{n=1}^{\infty}\int d^4y_n\cdots d^4y_1\,D_F(x,y_n)\nol \\
    && \ \ \ \ \ \ \ \ \ \ \ \ \ \ \ \ \ \times
    (-i m^2)\left[\prod_{j=1}^{n-1}D_F(y_{j+1},y_j)(-i m^2)\right]D_F(y_1,x')\,.\label{flatpert}
\eeqa
Substituting (\ref{momexp}) we have
\beqa
\Delta_F(x,x')    &=& -i\int\frac{d^4p}{(2\pi)^4}\frac{1}{p^2+M^2-i\epsilon}
    \sum_{n=0}^{\infty}\(-\frac{ m^2}{p^2+M^2-i\epsilon}\)^n e^{ip\cdot(x-x')}\nol \\
    &=& -i\int\frac{d^4p}{(2\pi)^4}\frac{1}{p^2+M^2+ m^2-i\epsilon}e^{ip\cdot(x-x')} \,,
\eeqa
which is the time-ordered two-point function for a free scalar field with mass $(M^2 +  m^2)^{1/2}$.

Now, let us apply the in-out perturbation theory to `our model' of conformally-coupled massive scalar field in de~Sitter expanding universe with the mass term treated as a perturbation.  This spacetime has the following metric:
\beq
    ds^2 = \frac{1}{H^2\lambda^2}\(-d\lambda^2+d\mathbf{x}\cdot d\mathbf{x}\) \,,
\eeq
where the conformal time $\lambda$ decreases from $\infty$ to $0$ towards the future, and where $H$ is the Hubble constant.
The Feynman two-point function for a conformally-coupled scalar field of mass $m$ of the two points $x=(\lambda,\mathbf{x})$ and $x'=(\lambda',\mathbf{x}')$ is~\cite{Schomblond:1976xc,Bunch:1978yq,Allen:1985wd}
\beq
\Delta_F(x,x') = \frac{H^2}{16\pi^2}\Gamma(a_+)\Gamma(a_{-})
F\left(a_+, a_-;2; Z(x,x')-i\epsilon\right)\,,  \label{exact}
\eeq
where
\beqa
Z(x,x') & \equiv & \frac{(\lambda+\lambda')^2 - \|\mathbf{x}-\mathbf{x}'\|^2}{4\lambda\lambda'}\,, \label{AHZ}\\
a_{\pm} & \equiv & \frac{3}{2} \pm \sqrt{\frac{1}{4} - \frac{ m^2}{H^2}}.
\eeqa
Here, $F(\alpha,\beta;\gamma;z)$ is Gauss's hypergeometric function~\cite{GR}.
The massless case reduces to
\beqa
D_F(x,x') & = & \frac{H^2}{16\pi^2}\frac{1}{1-Z(x,x')+i\epsilon}\\
&= & \frac{H^2}{4\pi^2} \frac{\lambda\lambda'}{\|\mathbf{x}-\mathbf{x}'\|^2-(\lambda-\lambda')^2+i\epsilon}\,.
\eeqa

In perturbative expansion
in $ m^2$ analogous to (\ref{flatpert}) the first-order term in the in-out Feynman two-point function is
\beqa
\Delta_F^{(1)}(x,x') & \equiv & \int\,d^4y\sqrt{-g(y)}D_F(x,y)(-i m^2)D_F(y,x') \nol \\
    & = & -\frac{i m^2}{(16\pi^2)^2}\int\frac{d\lambda_yd^3\mathbf{y}}{\lambda_y^4}
    \frac{1}{\left[1-Z(x,y)+i\epsilon\right]\left[1-Z(y,x')+i\epsilon\right]} \,, \label{first_order}
\eeqa
where we have let $y = (\lambda_y,\mathbf{y})$.
With $x$ and $x'$ fixed the integrand behaves like $\lambda_y^{-2}$ as $\lambda_y \to 0$.
Thus, the integral is infrared divergent due to its bad behaviour as $\lambda_y\to 0$, i.e.\ as point $y$ tends to future infinity.
However, since the exact time-ordered two-point function exists and is given by (\ref{exact}),
these infrared divergences do not imply that the time-ordered two-point function for massive scalar field is infrared divergent. It simply indicates that the in-out perturbation theory is inadequate here.
It will be found in section~\ref{sec:bogolubov2} that these divergences are due to the breakdown of perturbation theory in a term which should not be present in the two-point function in the first place.

\section{Bogolubov transformation between the true and free-theory vacua}\label{sec:bogolubov}

To understand the failure of the in-out perturbation theory in `our model' it is important to clarify the
relation between the true vacuum state and the free-theory vacuum state.  This relation is embodied in
the time-dependent Bogolubov transformation between these vacua, which we analyse in this section.

In the interaction picture the scalar field is expanded as a free field, i.e.\ a conformally-coupled
massless field, in `our model':
\beq
\phi_I(x) = \int \frac{d^3\mathbf{k}}{(2\pi)^3}
\left[ a_\mathbf{k} f_\mathbf{k}(x) + a^\dagger_\mathbf{k}f_\mathbf{k}^*(x)\right]\,,\label{AHphi1}
\eeq
where
\beq
    f_{\mathbf{k}}(x)
     = -\frac{iH\lambda}{\sqrt{2k}}e^{ik\lambda+i\mathbf{k}\cdot\mathbf{x}} \,. \label{AHfk}
\eeq
The free-theory vacuum state $|0,\mbox{free}\rangle$ satisfies $a_\mathbf{k}|0,\mbox{free}\rangle=0$ for all $\mathbf{k}$.
This field can also be expanded as
\beq
\phi_I(x) = \int \frac{d^3\mathbf{k}}{(2\pi)^3}
\left[ b_\mathbf{k}(\lambda) g_\mathbf{k}(x) + b_\mathbf{k}^\dagger(\lambda) g_\mathbf{k}^*(x)\right]\,,  \label{AHphi2}
\eeq
where $g_\mathbf{k}(x)$ are the exact mode functions for the scalar field with mass $m$:
\beq
    g_{\mathbf{k}}(x) = \frac{\sqrt{\pi}H}{2}\lambda^{3/2}H^{(1)}_{\nu}(k \lambda)e^{i\mathbf{k}\cdot\mathbf{x}}\,,
    \label{AHgk}
\eeq
with
\beq
    \nu = \sqrt{\frac{1}{4} - \frac{ m^2}{H^2}} \,.
\eeq
Here, $H_\nu^{(1)}(z)$ is the Hankel function of the first kind~\cite{GR}.
The operators $b_\mathbf{k}(\lambda)$ would be time-independent in the Heisenberg picture.  The true vacuum state $|\Omega(\lambda)\rangle$ is defined by requiring that $b_\mathbf{k}(\lambda)|\Omega(\lambda)\rangle = 0$ for all $\mathbf{k}$.  These annihilation and creation operators satisfy the standard commutation relations:
\beq
\left[ a_\mathbf{k},a_{{\mathbf{k}'}}^\dagger\right]
= \left[ b_\mathbf{k}(\lambda),b_{\mathbf{k}'}^\dagger(\lambda)\right]
= (2\pi)^3\delta^3(\mathbf{k}-\mathbf{k}')\,.
\eeq

We define the inner product of two functions, $f(x)$ and $g(x)$, which are not necessarily solutions to any wave equation, by
\beqa
(f,g)(\lambda) & \equiv & i\int d\Sigma n^\mu \left[ f^*(x)\partial_\mu g(x)- g(x)\partial_\mu f^*(x)\right]\label{AHinner} \\
& = & -\frac{i}{H^2\lambda^2}\int d^3\mathbf{x}\left[ f^*(x)\frac{\partial g(x)}{\partial \lambda} -
\frac{\partial f^*(x)}{\partial \lambda}g(x)\right] \,,
\eeqa
where $d\Sigma=d^3\mathbf{x}/(H\lambda)^3$ is the volume element on a hypersurface with $\lambda$ constant, and where
$n^\mu = -H\lambda(\partial/\partial \lambda)^\mu$ is the future-directed unit normal to this hypersurface.
Then we have
\beqa
    (f_{\mathbf{k}},f_{\mathbf{k}'})(\lambda) &=& (2\pi)^3\delta^3(\mathbf{k}-\mathbf{k}') \,, \label{ff1} \\
    (f^{\ast}_{\mathbf{k}},f^{\ast}_{\mathbf{k}'})(\lambda) &=& -(2\pi)^3\delta^3(\mathbf{k}-\mathbf{k}') \,, \label{ff2} \\
    (f_{\mathbf{k}},f^{\ast}_{\mathbf{k}'})(\lambda) &=& 0 \,, \label{ff3}
\eeqa
and similarly for $g_{\mathbf{k}}$, for all $\lambda$.
We can relate the modes $f_{\mathbf{k}}(x)$ and $g_{\mathbf{k}}(x)$ by a Bogolubov transformation:
\beq
    g_{\mathbf{k}}(x) = \int\frac{d^3\mathbf{k}'}{(2\pi)^3}\left[\alpha_{\mathbf{k}'\mathbf{k}}(\lambda)f_{\mathbf{k}'}(x)
        + \beta_{\mathbf{k}'\mathbf{k}}(\lambda)f^{\ast}_{\mathbf{k}'}(x)\right] \,. \label{g-f} 
\eeq
One can find these coefficients by the orthonormality property of the mode functions as
\beqa
    \alpha_{\mathbf{k}'\mathbf{k}}( \lambda) &=& (f_{\mathbf{k}'},g_{\mathbf{k}})  =
    \alpha_k( \lambda)(2\pi)^3\delta^3(\mathbf{k}-\mathbf{k}') \,, \label{alpha_diag}\\
    \beta_{\mathbf{k}'\mathbf{k}}( \lambda) &=& -(f^{\ast}_{\mathbf{k}'},g_{\mathbf{k}})= \beta_k( \lambda)(2\pi)^3\delta^3(\mathbf{k}+\mathbf{k}') \,, \label{beta_diag}
    \eeqa
where
\beqa
    \alpha_k( \lambda)
     &=& \frac{1}{2}\sqrt{\frac{\pi}{2k}}
        \[\(\frac{1}{2} \lambda^{-1/2} + ik \lambda^{1/2}\)H^{(1)}_{\nu}(k \lambda) + k \lambda^{1/2}H^{(1)\prime}_{\nu}(k \lambda)\] e^{-ik\lambda} \,, \label{alpha} \\
     \beta_k( \lambda) &= & \frac{1}{2}\sqrt{\frac{\pi}{2k}}
        \[\(\frac{1}{2} \lambda^{-1/2} - ik \lambda^{1/2}\)H^{(1)}_{\nu}(k \lambda) + k \lambda^{1/2}H^{(1)\prime}_{\nu}(k \lambda)\] e^{ik \lambda} \,. \label{beta}
\eeqa
Here the prime indicates the derivative with respect to the argument, i.e.\ $k\lambda$ in this case.  Thus, we have
\beq
g_\mathbf{k}(x) = \alpha_k( \lambda)f_\mathbf{k}(x) + \beta_k( \lambda)f_{-\mathbf{k}}^\ast(x)\,.\label{AHg-f}
\eeq
Then the orthonormality relations (\ref{ff1})-(\ref{beta_diag}) of the mode functions $f_\mathbf{k}(x)$ and $g_\mathbf{k}(x)$
imply
\beq
|\alpha_k(\lambda)|^2 - |\beta_k(\lambda)|^2 = 1\,. \label{AHhyperbolic}
\eeq
Let us recall that for large $k\lambda$ one has~\cite{GR}
\beq
H_\nu^{(1)}(k\lambda) \approx \sqrt{\frac{2}{\pi k\lambda}}\exp\left[i\left(k\lambda - \frac{1}{2}\pi \nu - \frac{\pi}{4}\right)\right]\,.
\eeq
Substituting this formula in (\ref{alpha}) and (\ref{beta}), one finds $\alpha_k(\lambda)\to 1$ and
$\beta_k(\lambda)\to 0$ as $k\lambda\to \infty$.
We can express $\alpha_k(\lambda)$ and $\beta_k(\lambda)$ in a form more suitable for perturbation in $m^2$
by differentiating them, using the differential equation satisfied by the Hankel
function $H_\mu^{(1)}(k\lambda)$ and then integrating with the boundary conditions at $k\lambda=\infty$.  The result is
\beqa
\alpha_k( \lambda) & = & 1 + \frac{1}{2}\sqrt{\frac{\pi}{2}}\,\frac{ m^2}{H^2}h_{-\,\nu}(k \lambda)\,,\\
\beta_k( \lambda) & = & \frac{1}{2}\sqrt{\frac{\pi}{2}}\,\frac{ m^2}{H^2}h_{+\,\nu}(k \lambda)\,, \label{AHbetak}
\eeqa
where
\beq
h_{\pm\,\nu}(z) \equiv \int_z^\infty \xi^{-3/2}e^{\pm i\xi}H_\nu^{(1)}(\xi)\,d\xi\,.
\eeq

Now, by equating the right-hand sides of (\ref{AHphi1}) and (\ref{AHphi2}), substituting (\ref{AHg-f}) and then comparing the coefficients of $f_\mathbf{k}(x)$ we find
\beq
    a_{\mathbf{k}} = \alpha_k( \lambda)b_{\mathbf{k}}( \lambda) + \beta^{\ast}_k( \lambda)b^{\dag}_{-\mathbf{k}}( \lambda) \,.
\eeq
This equation can be inverted using (\ref{AHhyperbolic}) as
\beq
b_\mathbf{k}(\lambda) = \alpha_k^\ast(\lambda) a_\mathbf{k} - \beta_k^\ast(\lambda) a_{-\mathbf{k}}^\dagger\,.
\eeq
These formulas characterize the relation between the true vacuum $|\Omega(\lambda)\rangle$ and
the free-theory vacuum $|0,{\rm free}\rangle$ because these vacua are defined by requiring
$b_\mathbf{k}(\lambda)|\Omega(\lambda)\rangle = a_\mathbf{k}|0,{\rm free}\rangle=0$ for all $\mathbf{k}$.
Since $\beta_k(\lambda)\to 0$ as $\lambda\to \infty$, the true and free-theory vacua agree at infinite past.
On the other hand we have $|\alpha_k(\lambda)|,\,\,|\beta_k(\lambda)|\to\infty$ as $\lambda \to 0$, i.e.\ in the infinite future.  This means that the true and free-theory vacua are orthogonal to each other in each momentum sector in the infinite future.  This feature makes the in-out two-point function different from the true two-point function as we shall see in the next section.  In the rest of this section we use the Bogolubov coefficients found here to calculate the rate of pair creation of free-theory particles.
(Since there is no energy conservation in de~Sitter expanding universe, it is not surprising that any interaction term in the Lagrangian leads to creation of free-theory particles on the free-theory in-vacuum-~\cite{Polyakov:2007mm,Higuchi:2008tn}.
However, the pair creation discussed here does not correspond to a physical phenomenon though it would be a physical process if the mass were changed from $0$ to $m$ only for a finite time.)

The number of free-theory particles in the true vacuum is
\beqa
N & = & \int \frac{d^3\mathbf{k}}{(2\pi)^3}
\langle \Omega( \lambda)|a_\mathbf{k}^\dagger a_\mathbf{k}|\Omega( \lambda)\rangle \nonumber \\
& = & \frac{V_c}{2\pi^2} \int_0^\infty dk\,k^2 |\beta_k( \lambda)|^2\,,
\eeqa
where $V_c = \int d^3\mathbf{x} = (2\pi)^3\delta^3(\mathbf{0})$ is the (infinite) \emph{coordinate} volume of the space.
Now, the physical wave number of the particle is given not by $k$ but by
$kH\lambda$ because $(H\lambda)^{-1}\|d\mathbf{x}\|$ is the physical length between $\mathbf{x}$ and $\mathbf{x}+d\mathbf{x}$.
Hence, $\kappa \equiv k\lambda$ is the physical wave number normalized by the Hubble constant $H$.
The free-theory particle number per unit physical volume $V(\lambda) = V_c/(H \lambda)^3$ at conformal time $\lambda$ can be expressed as
\beq
n  =  \frac{m^4}{16\pi H}\int_0^\infty d\kappa\,\kappa^2|h_{+\,\nu}(\kappa)|^2\,,
\eeq
where we have used (\ref{AHbetak}).
Thus, the number of free-theory particles in each physical momentum range $[\kappa,\kappa+d\kappa]$ in the true vacuum
$|\Omega(\lambda)\rangle$ is independent of the conformal time $\lambda$ as expected from de~Sitter invariance.

Now, a volume expanding with the universe whose coordinate volume is $V_c$ traces out a \emph{spacetime} volume
$V_c/(3H^4\lambda^3) = V(\lambda)/3H$.
This means that the number of free-theory particles created per unit spactime volume is $3H n$.  Hence, the rate of \emph{pair} creation per unit physical volume is $3Hn/2$, i.e.
\beq
\Gamma = \frac{3m^4}{32\pi}\int_0^\infty d\kappa\,\kappa^2 |h_{+\,\nu}(\kappa)|^2\,.
\eeq

To lowest order in $m^2$ we may approximate $h_{+\,\nu}(\kappa)$ by $h_{+\,1/2}(\kappa)$.  Thus we have
\beqa
    \Gamma
    &=& \frac{3m^4}{16\pi^2}\int_0^\infty\,d\kappa\kappa^2
    \int_{\kappa}^{\infty}\frac{e^{2i\xi}}{\xi^2}d\xi\int_{\kappa}^{\infty}\frac{e^{-2i\xi'}}{\xi^{\prime 2}}d\xi'\,.
\eeqa
We can integrate by parts twice to obtain
\beqa
    \Gamma &=& \frac{m^4}{8\pi^2}\mbox{Re}\int_0^{\infty}d\kappa\,\,\kappa e^{2i\kappa}
    \int_{\kappa}^{\infty}\frac{e^{-2i\xi}}{\xi^2}d\xi \nol \\
    &=& \frac{m^4}{8\pi^2}\mbox{Re}\,\int_0^\infty \left(\frac{\kappa e^{2i\kappa}}{2i}
    + \frac{e^{2i\kappa}-1}{4}\right)\frac{e^{-2i\kappa}}{\kappa^2}\,d\kappa \nol \\
    &=& \frac{m^4}{32\pi} \,. \label{AHpair}
\eeqa
This result will be reproduced by the standard perturbation theory in appendix~\ref{sec:emission}.

\section{in-out two-point function from Bogolubov transformation}\label{sec:bogolubov2}

In this section we write down the exact in-out two-point function in terms of the Bogolubov
coefficients in `our model' and show
that it is in fact infrared finite but does not equal the true two-point function.
Then we find that the divergence of the Bogolubov coefficients $\beta_k(\lambda)$ as $\lambda \to 0$
is the origin of the infrared divergences in perturbation theory found earlier in (\ref{first_order}).

We work in the Heisenberg picture in this section.  A conformally-coupled scalar
field $\phi(x)$ with mass $m$ is expanded as
\beq
\phi(x) = \int \frac{d^3\mathbf{k}}{(2\pi)^3}\left[ b_\mathbf{k}g_\mathbf{k}(x)+b_\mathbf{k}^\dagger g_\mathbf{k}^\ast(x)\right]\,,
\eeq
where $g_\mathbf{k}(x)$ are defined by (\ref{AHgk}).
The operators $b_\mathbf{k}$ are time-independent here because $\phi(x)$ satisfies the field equation with
mass $m$ in the Heisenberg picture.  The relation between the free-theory operators, which
are now time-dependent, and these operators is unchanged and given by
\beqa
a_\mathbf{k}(\lambda) & = & \alpha_k(\lambda)b_\mathbf{k} + \beta_k^*(\lambda)b_{-\mathbf{k}}^\dagger\,,\label{AHBk1}\\
b_\mathbf{k} & = & \alpha_k^\ast(\lambda)a_\mathbf{k}(\lambda) - \beta_k^\ast(\lambda)a_{-\mathbf{k}}^\dagger(\lambda)\,. \label{AHBk}
\eeqa
Let us define $a_\mathbf{k}^{\rm in} \equiv a_\mathbf{k}(\infty)$,
$\alpha_k^{\rm in}\equiv \alpha_k(\infty)$, $\beta_k^{\rm in} \equiv \beta_k(\infty)$ and
$a_\mathbf{k}^{\rm out} \equiv a_\mathbf{k}(0)$,
$\alpha_k^{\rm out} \equiv \alpha_k(0)$, $\beta_k^{\rm out}\equiv \beta_k(0)$.  Thus,
\beqa
a_\mathbf{k}^{\rm in} & = & \alpha_k^{\rm in}b_\mathbf{k} + \beta_k^{{\rm in}\,\ast}b_{-\mathbf{k}}^\dagger\,,
\label{AHin}\\
a_\mathbf{k}^{\rm out} & = & \alpha_k^{\rm out}b_\mathbf{k} + \beta_k^{{\rm out}\,\ast}b_{-\mathbf{k}}^\dagger\,.
\label{AHout}
\eeqa
We find from (\ref{AHBk1})-(\ref{AHout})
\beqa
    a_\mathbf{k}^{\rm out} & = & A_k a_\mathbf{k}^{\rm in} + B_k^* a_{-\mathbf{k}}^{{\rm in}\,\dagger}\,, \\
    a_\mathbf{k}^{\rm in} & = & A_k^* a_\mathbf{k}^{\rm out} - B_k^* a_{-\mathbf{k}}^{{\rm out}\,\dagger}\,,
\eeqa
where
\beqa
A_k & = & \alpha_k^{\rm out} \alpha_k^{{\rm in}\,\ast} - \beta_k^{{\rm out}\,\ast}\beta_k^{\rm in}\,,\\
B_k & = & \beta_k^{\rm out} \alpha_k^{{\rm in}\,\ast} - \alpha_k^{{\rm out}\,\ast}\beta_k^{\rm in}\,.
\eeqa

Now, we can write the in-out two-point function as
\beq
\langle \phi(x)\phi(x')\rangle_{\rm in-out}  =   \frac{\langle 0,{\rm out}|\phi(x)\phi(x')|0,{\rm in}\rangle}
{\langle 0,{\rm out}|0,{\rm in}\rangle}\,.
\eeq
(We have dropped time ordering here because it is not essential.)
Since the state $\phi(x')|0,{\rm in}\rangle$ and $\phi(x)|0,{\rm out}\rangle$
are one-particle states, the numerator can be written
\beqa
\langle 0,{\rm out}|\phi(x)\phi(x')|0,{\rm in}\rangle
& =  & \int \frac{d^3\mathbf{k}}{(2\pi)^3} \frac{d^3\mathbf{k}'}{(2\pi)^3}
\langle 0,{\rm out}|\phi(x)a_\mathbf{k}^{{\rm out}\,\dagger}|0,{\rm out}\rangle \nonumber \\
&& \times \langle 0,{\rm out}|a_\mathbf{k}^{\rm out} a_{\mathbf{k}'}^{{\rm in}\,\dagger}|0,{\rm in}\rangle
\langle 0,{\rm in}|a_{\mathbf{k}'}^{{\rm in}}\phi(x')|0,{\rm in}\rangle\,.
\eeqa
We readily find
\beqa
\langle 0,{\rm out}|\phi(x)a_\mathbf{k}^{{\rm out}\,\dagger}|0,{\rm out}\rangle
& = & \alpha_k^{{\rm out}\,\ast}g_\mathbf{k}(x) - \beta_k^{\rm out}
g_{-\mathbf{k}}^\ast(x)\,,\\
\langle 0,{\rm in}|a_{\mathbf{k}'}^{{\rm in}}\phi(x')|0,{\rm in}\rangle
& = & \alpha_k^{{\rm in}}g_{\mathbf{k}'}^\ast(x') - \beta^{{\rm in}\,\ast}_kg_{-\mathbf{k}'}(x')\,,\\
\langle 0,{\rm out}|a_\mathbf{k}^{\rm out}a_{\mathbf{k}'}^{{\rm in}\,\dagger}|0,{\rm in}\rangle
& = & (A_k^\ast)^{-1}(2\pi)^3\delta^3(\mathbf{k}-\mathbf{k'})\langle 0,{\rm out}|0,{\rm in}\rangle\,.
\eeqa
Thus, we obtain
\beq
\langle \phi(x)\phi(x')\rangle_{\rm in-out} =
\int \frac{d^3\mathbf{k}}{(2\pi)^3}
(1-\gamma_k^{\rm out}\gamma_k^{{\rm in}\,*})^{-1}\left[g_\mathbf{k}(x)-\gamma_k^{\rm out}g_{-\mathbf{k}}^*(x)\right]
\left[g_\mathbf{k}^*(x') - \gamma_k^{{\rm in}\,*}g_{-\mathbf{k}}(x')\right]\,, \label{genBog}
\eeq
where $\gamma_k^{\rm in} \equiv \beta_k^{\rm in}/\alpha_k^{{\rm in}\,*}$
and $\gamma^{\rm out}\equiv \beta_k^{\rm out}/\alpha_k^{{\rm out}\,*}$.
In our case $\beta_k^{\rm in} = 0$. That is,  $|\Omega\rangle = |0,{\rm in}\rangle$.  Hence
\beq
\langle \phi(x)\phi(x')\rangle_{\rm in-out} =
\int \frac{d^3\mathbf{k}}{(2\pi)^3}
\left[ g_\mathbf{k}(x)g_{\mathbf{k}}^*(x')
- \gamma_k^{\rm out}g_{-\mathbf{k}}^*(x)g_\mathbf{k}^*(x')\right]\,. \label{AHdiverge}
\eeq
The first term gives the correct two-point function by itself, and hence the in-out two-point function
does not agree with the correct two-point function if $\beta_k^{\rm out}\neq 0$ for some $k$.  Nevertheless,
since $|\gamma_k^{\rm out}|\leq 1$ in general --- in fact $|\gamma_k^{\rm out}| =1$ in our case ---
the in-out two-point function is not infrared divergent.  This might appear to contradict the
divergence in (\ref{first_order}).
However, by formally working to order $m^2$ we recover the divergent term
we have encountered in perturbation theory as follows.  To lowest order in $ m^2$ we have
$\gamma_k^{\rm out} = \beta_k^{\rm out} + o(m^2)$.  Hence, to lowest order we have
\beq
\gamma_k^{\rm out} = - \frac{i m^2}{2H^2}\int\frac{e^{2i\xi}}{\xi^2}d\xi\,,  \label{AHgamma}
\eeq
which is infrared divergent.
Substituting (\ref{AHgamma}) in the second term of (\ref{AHdiverge}) and approximating the functions $g_\mathbf{k}(x)$ by
$f_\mathbf{k}(x)$ given by (\ref{AHfk}), one can express this term as
\beq
\langle \phi(x)\phi(x')\rangle_{\rm in-out}^{\rm div}
= - \frac{1}{2}i m^2\lambda\lambda'\int \frac{d^3\mathbf{k}}{(2\pi)^32k}
e^{i\mathbf{k}\cdot(\mathbf{x}-\mathbf{x}')-ik(\lambda+\lambda')}
\int \frac{e^{2i\xi}}{\xi^2}d\xi\,.  \label{AHmomdiv}
\eeq
One can show that this formula reproduces the infrared divergences found earlier by using (\ref{first_order})
in the momentum expansion of the free-theory Feynman two-point function,
\beq
D_F(x,x') = H^2\theta(\lambda'-\lambda)\lambda\lambda' \int \frac{d^3\mathbf{k}}{(2\pi)^32k}
e^{ik(\lambda-\lambda') + i\mathbf{k}\cdot(\mathbf{x}-\mathbf{x}')}
+ (\lambda\leftrightarrow \lambda')
\,.  \label{AHmomexp}
\eeq
Thus, as
we stated before, the infrared divergences in the in-out Feynman two-point function in `our model' are
due to breakdown of perturbation theory in a term that should not be present, i.e.\ the second term in (\ref{AHdiverge}).

\section{Euclidean approach}\label{sec:Euclidean}

As we found in the previous section,
the discrepancy between the true vacuum state and the free-theory out-vacuum state invalidates the use of the in-out formalism for constructing the two-point function of massive scalar field from massless scalar field.  In this section we point out that the Euclidean approach does work.  This means that perturbation theory in
thermal field theory inside the cosmological horizon with the Gibbons-Hawking temperature $H/2\pi$ correctly gives the two-point function in `our model'.

The static metric of de~Sitter spacetime inside the cosmological horizon is
\beq
ds^2 = -(1-H^2r^2)dt^2 +(1-H^2r^2)^{-1}dr^2 + r^2(d\theta^2 + \sin^2\theta\,d\varphi^2)\,.
\eeq
By letting $t=i\tau$ and $Hr=\sin\chi$, we obtain
\beq
ds^2 = H^{-2}\left[ \cos^2\chi\,d(H\tau)^2 + d\chi^2 + \sin^2\chi\left(d\theta^2 + \sin^2\theta\,d\varphi^2\right)\right],
\eeq
which is the metric of $S^4$ of radius $H^{-1}$ if $\tau$ is periodically identified with period $2\pi/H$.  Thus, the field theory on $S^4$ of radius $H^{-1}$ describes the thermal field theory inside the cosmological horizon with temperature $H/2\pi$.

Let the full set of mode functions on $S^4$ of radius $H^{-1}$ be
$\phi_{\ell \sigma}(x)$, where
\beq
-\Box \phi_{\ell \sigma}(x) = \ell(\ell+3)H^2\phi_{\ell\sigma}(x)\,,\,\,\ell=0,1,2,\ldots.
\eeq
The label $\sigma$ differentiates the modes with the same quantum number $\ell$.  Let $\phi_{\ell\sigma}(x)$ be
orthonormal:
\beq
\int d^4x\sqrt{g(x)}\,\phi^*_{\ell \sigma}(x)\phi_{\ell'\sigma'}(x) =
\delta_{\ell\ell'}\delta_{\sigma\sigma'}\,.
\eeq
Then, the Green's function of the unperturbed theory with the equation $(-\Box + 2H^2)\phi(x)=0$ is
\beq
D_E(x,x')  =  \sum_{\ell=0}^\infty
\sum_{\sigma}\frac{\phi_{\ell\sigma}(x)\phi_{\ell\sigma}^*(x')}{\ell(\ell+3)H^2+2H^2}\,,
\eeq
which satisfies
\beq
(-\Box_x + 2H^2)D_E(x,x') = \frac{1}{\sqrt{g(x)}}\delta^4(x,x')\,.
\eeq
The Green's function for the theory with the equation $(-\Box + 2H^2 +  m^2)\phi(x) =0$ can be obtained perturbatively as
\beqa
    \Delta_E(x,x')
    & = & D_E(x,x')
    + \sum_{n=1}^{\infty}\int d^4y_n\cdots d^4y_1\,D_E(x,y_n)\nol \\
    && \ \ \ \ \ \ \ \ \ \ \ \ \ \ \ \ \ \times
    (- m^2)\left[\prod_{j=1}^{n-1}D_E(y_{j+1},y_j)(- m^2)\right]D_E(y_1,x')\nonumber \\
&  = & \sum_{\ell=0}^\infty
\sum_{\sigma}\frac{\phi_{\ell\sigma}(x)\phi_{\ell\sigma}^*(x')}{\ell(\ell+3)H^2+2H^2+
m^2}\,,
\eeqa
which is the correct Green's function, from which the time-ordered
two-point function in the Euclidean vacuum is obtained by
analytic continuation.


\section{The two-point function in the in-in formalism}\label{sec:in-in}

Since the Bogolubov transformation becomes trivial for each $\mathbf{k}$ as $\lambda\to \infty$, i.e.\
$\beta_k^{\rm in} = 0$, we have $|0,{\rm in}\rangle = |\Omega\rangle$.  Now, the two-point Wightman function in the in-in
formalism is by definition
\beq
\Delta_{\rm in}(x,x') = \langle 0,\mbox{in}|\phi(x)\phi(x')|0,\mbox{in}\rangle\,,
\eeq
where $\phi(x)$ is the Heisenberg operator satisfying the field equation
$(-\Box + 2H^2 + m^2)\phi(x) = 0$.  This two-point function clearly agrees with the true two-point function
$\langle \Omega|\phi(x)\phi(x')|\Omega\rangle$ because $|0,{\rm in}\rangle = |\Omega\rangle$.  Thus, the in-in perturbation theory works
in `our model'.  In the rest of this section we verify this fact to order $m^2$ by a concrete calculation.

We define the interaction Hamiltonian in terms of the free field, $\phi_I(x)$, i.e.\ the field in the interaction picture, satisfying $(-\Box + 2H^2)\phi_I(x) = 0$ as
\beq
H_I( \lambda) = \frac{ m^2}{2}\int \frac{d^3\mathbf{x}}{(H \lambda)^3}:\phi_I^2(x):\,, \label{AHHI}
\eeq
where $:\cdots:$ denotes normal ordering.  Recall that $d^3\mathbf{x}/(H \lambda)^{3}$ is the induced volume element on the hypersurface with $\lambda$ constant.
Then the Heisenberg operator $\phi(x)$ is related to the field $\phi_I(x)$ as
\beq
\phi(x) =
\left\{\hat{T}\exp\(i\int_\lambda^{\infty}\frac{dv}{Hv}\,H_I(v)\)\right\}
    \phi_I(x)\left\{T\exp\(-i\int^{\infty}_\lambda\frac{dt}{Hv}\,H_I(v)\)\right\}\,,\label{AHin-in}
\eeq
where $T$ $(\hat{T})$ indicates (anti-)time ordering.
This formula is more conveniently written (see, e.g. \cite{Weinberg:2005vy}) as
\beq
\phi(x) = \sum_{N=0}^\infty \varphi^{(N)}(x)\,,  \label{AHphiexp}
\eeq
where
\beqa
\varphi^{(0)}(x) & = & \phi_I(x)\,, \label{AHfirsteq}\\
\varphi^{(N)}(x) & = & i\int_\lambda^\infty \frac{dv}{Hv}\left[H_I(v),\varphi^{(N-1)}(v,\mathbf{x})\right]\,,\,\,
N\geq 1\,. \label{AHrecursion}
\eeqa
Then, by substituting (\ref{AHHI}) in (\ref{AHrecursion}) one can show by induction
\beq
\varphi^{(N)}(x) = - m^2\int d^4x'\sqrt{-g(x')}\,G_R(x,x')\varphi^{(N-1)}(x')\,, \label{AHGreeneq}
\eeq
where the retarded Green's function $G_R(x,x')$ is given by
\beq
    G_R(x,x') \equiv i\theta( \lambda'- \lambda)\langle0,\mbox{in}|[\phi_I( \lambda,\mathbf{x}),\phi_I( \lambda',\mathbf{x}')]|0,\mbox{in}\rangle
    \,. \label{G_R}
\eeq
(The state $|0,\mbox{in}\rangle$ in this equation can be replaced by
any other state since the commutator $[\phi_I( \lambda,\mathbf{x}),\phi_I( \lambda',\mathbf{x}')]$ is a c-number.)
This function is given explicitly as (see, e.g.\ \cite{Allen:1985wd,Higuchi:2008fu})
\beq
    G_R(x,x') =
    \theta( \lambda'- \lambda)\frac{H^2\lambda \lambda'}{4\pi\|\textbf{x}-\textbf{x}'\|} \delta( \lambda'- \lambda-\|\textbf{x}-\textbf{x}'\|) \,. \label{Retarded_Green}
\eeq
The Wightman two-point function to order $ m^2$ in the in-in formalism is
\beq
\Delta^{(1)}(x,x') = D(x,x') + \Phi(x,x') + \Phi^*(x',x)\,,
\eeq
where
\beqa
D(x,x') & = & \langle 0,{\rm in}|\phi_I(x)\phi_I(x')|0,{\rm in}\rangle \nonumber \\
& = & \frac{H^2}{4\pi^2}\frac{\lambda\lambda'}{\|\mathbf{x}-\mathbf{x}'\|^2 - (\lambda'-\lambda-i\epsilon)^2}\nonumber \\
& = & \frac{H^2}{16\pi^2}\frac{1}{1-Z(x,x')+i\epsilon\,\mbox{sgn}(\lambda'-\lambda)}
\eeqa
is the free-theory two-point function. The function $\Phi(x,x')$ is defined by
\beq
\Phi(x,x')  \equiv  \langle 0,\mbox{in}|\phi_I(x)\varphi^{(1)}(x')|0,\mbox{in}\rangle\,, \label{AHPhidef}
\eeq
where $\varphi^{(1)}(x)$ is given by (\ref{AHGreeneq}).

Now, if $\Delta_{m^2}(x,x')$ is the Wightman two-point function for the
conformally-coupled scalar field with mass $m$, then (see, e.g.\ \cite{AllenTuryn,Higuchi:2001uv})
\beq
\left. \frac{\partial\ }{\partial  m^2}\Delta_{m^2}(x,x')\right|_{m^2=0}
= \frac{1}{16\pi^2Z(x,x')}\log\left[1-Z(x,x')+i\epsilon\,\mbox{sgn}(\lambda'-\lambda)\right]\,.
\eeq
Therefore, what we need to show is
\beq
\Phi(x,x')+\Phi^*(x',x) =\frac{ m^2}{16\pi^2}\frac{1}{Z(x,x')}\log\left[1-Z(x,x')+i\epsilon\,\mbox{sgn}(\lambda'-\lambda)\right]\,.
\label{AHeqtoshow}
\eeq
We find, using (\ref{AHPhidef}), (\ref{AHGreeneq}) and (\ref{AHfirsteq}),
\beq
    \Phi(x,x') = - m^2\int\,d^4y\sqrt{-g(y)}G_R(x',y)D(x,y)\,.
\eeq
We have, after redefining the integration variables $\mathbf{y}$ as $\mathbf{y}+\mathbf{x}'$,
\beq
\Phi(x,x')        = \frac{ m^2\lambda\lambda'}{16\pi^3}\int \frac{d^3\mathbf{y}}{y(y+\lambda')^2}
\frac{1}{(\lambda'-\lambda+y-i\epsilon)^2-\|\mathbf{x}-\mathbf{x}'-\mathbf{y}\|^2}\,,
\eeq
where $y\equiv\|\mathbf{y}\|$.
Performing the angle integral and then the integral over $y$, we find
\beqa
\Phi(x,x') & = & \frac{ m^2 \lambda\lambda'}{8\pi^2 \|\mathbf{x}-\mathbf{x}'\|}
\left[ \frac{1}{\lambda'+\lambda+i\epsilon-\|\mathbf{x}-\mathbf{x}'\|}
\log \frac{\lambda'-\lambda-i\epsilon + \|\mathbf{x}-\mathbf{x}'\|}{2\lambda'}\right. \nonumber \\
&& \ \ \ \ \ \ \ \ \ \
\left. - \frac{1}{\lambda+\lambda' +i\epsilon + \|\mathbf{x}-\mathbf{x}'\|}\log
\frac{\lambda'-\lambda-i\epsilon - \|\mathbf{x}'-\mathbf{x}\|}{2\lambda'}\right]\,.
\eeqa
Since this function is not singular at $\lambda'+\lambda = \pm\|\mathbf{x}-\mathbf{x}'\|$, we may
write $\lambda+\lambda'+i\epsilon$ simply as $\lambda+\lambda'$ without any ambiguity.  Then we obtain
\beq
\Phi(x,x')+\Phi^*(x',x)  =  \frac{ m^2}{16\pi^2}\frac{4\lambda\lambda'}{(\lambda'+\lambda)^2- \|\mathbf{x}-\mathbf{x}'\|^2}
\log \frac{\|\mathbf{x}-\mathbf{x}'\|^2 - (\lambda'-\lambda-i\epsilon)^2}{4\lambda\lambda'}\,,
\eeq
which can be shown to equal (\ref{AHeqtoshow}) using the definition (\ref{AHZ}) of $Z(x,x')$.

\section{Summary}\label{sec:remarks}

In this paper we clarified why the in-out perturbation theory
fails in de~Sitter expanding universe in `our model' of conformally-coupled
massive scalar field theory with the mass term treated as a perturbation.  We also pointed out that
the Euclidean and in-in formulations
correctly reproduce the exact theory in `our model'.
These results
support the view that properties of interacting field theory in de~Sitter expanding universe are more reliably extracted using either the Euclidean or in-in formalism rather than the in-out formalism. Any results using the in-out scattering theory, e.g.~the decay rate of a scalar particle due to self-interaction~\cite{Boyanovsky:2004gq,Boyanovsky:2004ph}, may need to be re-examined by taking into account the difference between the free-theory in- and out-vacua in interacting field theories.

\acknowledgments

We thank Don Marolf for useful correspondence, which motivated this work and Benard Kay for useful comments.
One of the authors (A.H.) thanks the Astro-Particle Theory and Cosmology Group and the Department of Applied Mathematics at University of Sheffield, where part of this work was carried out, for kind hospitality.

\appendix

\section{Apparent pair creation in the standard perturbation theory}\label{sec:emission}

In this appendix we derive the pair-creation rate (\ref{AHpair}) of the free-theory particles using the standard perturbation theory in the interaction picture.  A similar calculation has been presented for $\varphi^4$ theory in \cite{Higuchi:2008tn}, which demonstrated
that the free-theory vacuum evolves by emission of free-theory particles in the interaction picture.

We define the transition amplitude $\mathcal{A}(\mathbf{k}_1,\mathbf{k}_2)$ from the free-theory in-vacuum state,
$|0,\mbox{free}\rangle$,
to a state with two free-theory particles,
$|\mathbf{k}_1,\mathbf{k}_2\rangle=a^{\dag}_{\mathbf{k}_1}a^{\dag}_{\mathbf{k}_2}|0,\mbox{free}\rangle$, as
\beqa
    \mathcal{A}(\mathbf{k}_1,\mathbf{k}_2)
        &=& \int \frac{d\lambda}{H\lambda}\langle\mathbf{k}_1\mathbf{k}_2|H_I|0,\mbox{free}\rangle \nol \\
        &=& \frac{ m^2}{2}\int_{-\infty}^{\infty}\,dt\,\frac{e^{Ht}}{\sqrt{k_1k_2}}\exp\left[-\frac{i}{H}(k_1+k_2)e^{-Ht}\right]
        (2\pi)^3\delta^3(\mathbf{k}_1+\mathbf{k}_2) \,,
\eeqa
where we have
made the change of variable from $\lambda$ to $t=-H^{-1}\log H\lambda$, which is the proper time of the timelike geodesic with $\mathbf{x}$ constant.
The interaction Hamiltonian $H_I$ is given by (\ref{AHHI}).
Then the transition probability is
\beqa
    \mathcal{P} &=& \frac{1}{2!}\int\frac{d^3\mathbf{k}_1}{(2\pi)^3}\frac{d^3\mathbf{k}_2}{(2\pi)^3}
        |\mathcal{A}(\mathbf{k}_1,\mathbf{k}_2)|^2 \nol \\
& = &     \frac{m^4}{8}
        \int\frac{d^3\mathbf{k}_1d^3\mathbf{k}_2}{(2\pi)^3}
        \int_{-\infty}^{\infty}\,dt_1\int_{-\infty}^{\infty}\,dt_2
        \frac{e^{H(t_1+t_2)}}{k_1k_2}
        \exp\[-\frac{i(k_1+k_2)}{H}\(e^{-Ht_1}-e^{-Ht_2}\)\] \nol \\
        && \ \ \ \ \ \ \ \ \ \ \ \ \ \ \ \times\delta^3(\mathbf{k}_1+\mathbf{k}_2)V_c \,,
\eeqa
where $V_c=\int_{-\infty}^{\infty}d^3\mathbf{x}=(2\pi)^3\delta^3(\mathbf{0})$ is the infinite coordinate volume.
Changing the variables again as $T=(t_1+t_2)/2$ and $\tau=t_1-t_2$ and integrating over $\mathbf{k}_2$, we find
\beqa
    \mathcal{P} & = & \frac{m^4}{16\pi^2}
        \int_0^\infty dk_1
        \int_{-\infty}^{\infty}\,dT\int_{-\infty}^{\infty}\,d\tau\, e^{2HT}
        \exp\[\frac{4ik_1}{H}e^{-HT}\sinh\frac{H\tau}{2}\]V_c \nonumber \\
& = & \frac{m^4}{32\pi^2} \int_{-\infty}^{\infty}dT
        \int_0^{\infty}d\kappa\int_{-\infty}^{\infty}\,d\xi\,\,\exp\(i\kappa\sinh\xi\)
V_c\,e^{3HT} \,,
\eeqa
where we have made the change of variables $\kappa=(4k_1/H)e^{-HT}$ and $\xi=H\tau/2$.
Since $V_c e^{3HT}$ can be interpreted as the physical volume of the space at time $T$, we conclude that
the pair-creation rate $\Gamma$ per unit physical volume is
\beqa
    \Gamma & = & \frac{m^4}{32\pi^2}
        \int_0^{\infty}d\kappa\int_{-\infty}^{\infty}\,d\xi\,\,\exp\(i\kappa\sinh\xi\)\nonumber \\
& = &     \frac{m^4}{32\pi} \,, \label{Emission_Rate}
\eeqa
which is in agreement with (\ref{AHpair}).

\section{In-out perturbation theory about the wrong mass in Minkowski spacetime}\label{A}

In this appendix we discuss free scalar field $\phi$ of mass $(M^2 + m^2)^{1/2}$ in Minkowski spacetime with the term
$-\frac{1}{2}m^2\phi^2$ in the Lagrangian treated as a perturbation.  Thus, the mode functions for the `free' theory, $f_\mathbf{k}(x)$, and those for the exact theory,
$g_\mathbf{k}(x)$, are
\beqa
f_\mathbf{k}(x) & = & \frac{1}{\sqrt{2k_0}}e^{-ik_0t}e^{i\mathbf{k}\cdot\mathbf{x}}\,,\\
g_\mathbf{k}(x) & = & \frac{1}{\sqrt{2K_0}}e^{-iK_0t}e^{i\mathbf{k}\cdot \mathbf{x}}\,,
\eeqa
where $k_0 \equiv (\mathbf{k}^2 + M^2)^{1/2}$ and $K_0 \equiv (\mathbf{k}^2+M^2 + m^2)^{1/2}$.
The Bogolubov coefficients can be found from (\ref{alpha_diag}) and (\ref{beta_diag}) (with $\lambda$ replaced by $t$)
with the inner product (\ref{AHinner}) adapted to
Minkowski spacetime. The result is
\beqa
\alpha_k(t) & = & \frac{1}{2}\left[ \left(\frac{k_0}{K_0}\right)^{1/2} + \left(\frac{K_0}{k_0}\right)^{1/2}\right]
e^{i(k_0-K_0)t}\,,\\
\beta_k(t) & = & \frac{1}{2}\left[ \left(\frac{k_0}{K_0}\right)^{1/2} - \left(\frac{K_0}{k_0}\right)^{1/2}\right]
e^{-i(k_0+K_0)t}\,.
\eeqa
We define $\alpha_k^{\rm in} \equiv \alpha_k(-T)$, $\beta_k^{\rm in} \equiv \beta_k(-T)$ and
$\alpha_k^{\rm out} \equiv \alpha_k(T)$,
 $\beta_k^{\rm out} \equiv \beta_k(T)$, and let
$T \to \infty(1-i\epsilon)$ at the end following the standard procedure (see, e.g. \cite{Peskin:1995ev}).
Then
\beq
\gamma_k^{\rm out} = \gamma_k^{{\rm in}\,*} =  - \kappa e^{-2iK_0 T}\,,
\eeq
where
\beq
\kappa \equiv \frac{K_0 - k_0}{K_0+k_0}\,.
\eeq
Thus, the in-out two-point function given by (\ref{genBog}) reads (for $t>t'$)
\beq
\langle \phi(x)\phi(x')\rangle_{\rm in-out} =
\int \frac{d^3\mathbf{k}}{(2\pi)^3}
\frac{\left[g_\mathbf{k}(x)+\kappa e^{-2iK_0T}g_{-\mathbf{k}}^*(x)\right]
\left[g_\mathbf{k}^*(x') + \kappa e^{-2iK_0T}g_{-\mathbf{k}}(x')\right]}
{1- \kappa^2 e^{-4iK_0T}}\,.
\eeq
Since $e^{-2iK_0T} \to 0$ in the limit $T\to \infty (1-i\epsilon)$,  we obtain
\beq
\langle \phi(x)\phi(x')\rangle_{\rm in-out} =  \int \frac{d^3\mathbf{k}}{(2\pi)^3}
g_\mathbf{k}(x) g_\mathbf{k}^*(x')\,,
\eeq
which is the exact two-point function for the scalar field with mass $(M^2 + m^2)^{1/2}$ (for $t > t'$).

\end{document}